\documentclass[double]{article}
\usepackage{times}
\usepackage{algorithm}
\usepackage{algorithmic}
\usepackage{wrapfig}
\usepackage{verbatim}
\usepackage{amsmath}
\usepackage{graphicx}
\usepackage{subcaption}
\usepackage[active]{srcltx}
\usepackage{color}
\usepackage{lineno}
\usepackage{dirtytalk}
\usepackage{amsthm}
\usepackage{authblk}
\usepackage{listings}
\usepackage{tikz}
\usepackage{longtable}
\usetikzlibrary{shapes.geometric, arrows}

\usepackage{epsfig,graphicx,amsfonts}
\usepackage{amsmath,amsthm,amssymb}
\usepackage{xcolor}

\usepackage{adjustbox}
\usepackage{multirow}
\usepackage{setspace}
\usepackage{hyperref}
\usepackage{comment}

\long\def\comment #1\commentend{}

\begin{document}

\title{\Large Transforming Norm-based To Graph-based Spatial Representation for Spatio-Temporal Epidemiological Models}

\author{Teddy Lazebnik$^{1,2,*}$\\
\(^1\) Department of Information Systems, University of Haifa, Haifa, Israel \\ \(^2\) Department of Computing, Jonkoping University, Jonkoping, Sweden \\ 
\(^*\) Corresponding author: lazebnik.teddy@gmail.com \\

}

\date{ }

\maketitle 

\begin{abstract}
\noindent
Pandemics, with their profound societal and economic impacts, pose significant threats to global health, mortality rates, economic stability, and political landscapes. In response to these challenges, numerous studies have employed spatio-temporal models to enhance our understanding and management of these complex phenomena. These spatio-temporal models can be roughly divided into two main spatial categories: norm-based and graph-based. Norm-based models are usually more accurate and easier to model but are more computationally intensive and require more data to fit. On the other hand, graph-based models are less accurate and harder to model but are less computationally intensive and require fewer data to fit. As such, ideally, one would like to use a graph-based model while preserving the representation accuracy obtained by the norm-based model. In this study, we explore the ability to transform from norm-based to graph-based spatial representation for these models. We first show no analytical mapping between the two exists, requiring one to use approximation numerical methods instead. We introduce a novel framework for this task together with twelve possible implementations using a wide range of heuristic optimization approaches. Our findings show that by leveraging agent-based simulations and heuristic algorithms for the graph node's location and population's spatial walk dynamics approximation one can use graph-based spatial representation without losing much of the model's accuracy and expressiveness. We investigate our framework for three real-world cases, achieving 94\% accuracy preservation, on average. Moreover, an analysis of synthetic cases shows the proposed framework is relatively robust for changes in both spatial and temporal properties. \\ 

\end{abstract}

\maketitle \thispagestyle{empty}
\pagestyle{myheadings} \markboth{Draft:  \today}{Draft:  \today}
\setcounter{page}{1}

\section{Introduction}
\label{sec:introduction}
Pandemics over history caused intense impact on societies across centuries, leaving an indelible mark on mortality rates, economic stability, and political landscapes \cite{pandemic_important}. In the last decades, the persistent threat of emerging and reemerging pandemics and epidemics looms large over humanity \cite{pandemic_duration}. Indeed, recent outbreaks such as the severe acute respiratory syndrome (SARS) in 2003, the H1N1 influenza pandemic in 2009, the Middle East respiratory syndrome coronavirus (MERS-CoV) in 2012, the Ebola virus in 2014, and the ongoing global severe acute respiratory syndrome coronavirus 2 (SARS-CoV-2) pandemic, underscore the urgency of comprehending, preparing, and managing pandemics \cite{who_problem,intro_pandemic_example_1,teddy_labib_financing_pandemic,intro_pandemic_example_2,example_influenza}.

To this end, multiple mathematical and computational models have been proposed to ameliorate our understanding of the pandemic spread and advance our ability to control it \cite{social_distance_1,sir_clinical,review_zika,review_covid}. These models can be roughly divided into two primary groups: statistical models and mechanistic models. The former group employs a data-driven approach without making explicit assumptions about the underlying dynamics, while the latter relies on theoretical principles to elucidate the mechanisms governing pandemic dynamics. Statistical models, on the other hand, commonly leverage statistical and machine learning (ML) methods to project a diverse array of outcomes. Within this category, a multitude of models has been developed utilizing various techniques, including auto-regressive time series methods, Bayesian optimization techniques, and, more recently, deep learning models \cite{stat_model_1, different_approach_from_sir, different_approach_from_sir_2, stat_model_6, stat_model_5}. These models have been shown to provide promising results on the pandemic they trained on but they usually do not generalize well on other cases, require large amounts of data to obtain reasonable performance, and are very sensitive to noise. The latter group, exemplified by the SIR model, operates based on a predefined set of rules or processes governing interactions among individuals within a population \cite{models_good_1, models_good_2, models_good_3, miller2017mathematical}. At the forefront of this approach stands the Susceptible-Infected-Recovered (SIR) model \cite{sir_popular_5, sir_popular_4, infection_graph_main}, originally proposed by Kermack and McKendrick \cite{first_sir}. Models belonging to this group require more effort from the modeler as one needs to manually detect, define, and formalize the main interactions and processes occurring in a pandemic for the model to capture the dynamics correctly. That said, once done efficiently, models from this group are more generalizable, adaptable, explainable, and require less data. 

Focusing on the mechanistic models, a growing number of works adopted a spatiotemporal model formalization as the spatial nature of pandemics is repeatedly shown to take a significant part in the dynamics \cite{optimization_method_1,optimization_method_1b,teddy_economic}. Intuitively, the first SIR model \cite{first_sir} assumes a well-mixture property of the population which means that for each point in time \(t\), the probability two random individuals in the population would interact is uniformly distributed. This assumption is problematic as it is known not to be accurate for large-scale scenarios such as cities or countries \cite{spatio_1,spatio_2,spatio_3}. Commonly, modelers mistakenly assume one can use the well-mixture assumption for short periods of time or small spatial locations, such as rooms or buildings, but recent studies show that even for these cases the approximation is sub-optimal if one desires reasonable accurate predictions \cite{b1,b5,b2,b3}. Therefore, spatial dynamics are important in order to capture pandemic spread. One can categorize the spatial component into two main groups: Graph-based and Norm-based. The graph-based approach assumes a graph, \(G = (V, E)\) such that \(E \subseteq V \times V\), where \(V\) are the nodes and \(E\) are the edges between them. Interestingly, the nodes (\(V\)) can either represent locations where individuals are located at in a given time \cite{spatial_graph_intro,spatial_graph_1,spatial_graph_1a} or the individuals themselves. For the first case, the edges are usually the possible connections between the physical location and the individuals performing some kind of walk between the nodes \cite{infection_graph_main,spatial_graph_1}. This approach abstracts the physical domain of the pandemic in favor of a more abstract and computationally attractive representation allowing the modeler to focus on \say{conceptual} locations, often ignoring the transformation between them \cite{alexi}. On the other hand, for the second case, the edges indicate the interaction between the individuals, often associated with infectious routs \cite{norm_based_1,norm_based_2}. In a complementary manner, the norm-based approach assumes a continuous space, rather than a discrete one enforced in the graph-based approach, where individuals are moving and interacting over time. This setting is usually more computationally expensive while providing a more physically accurate presentation of reality \cite{numerical_software_compare}. 

One can easily notice that the norm-based representation is more expressive than the graph-based one. Nonetheless, this expressiveness can be computationally overwhelming while relatively easy to model and comprehend compared to the more abstract graph-based approach. Therefore, it is interesting to pose the question of whether we can transform models from norm-based to graph-based representations, in a way that preserves the model's accuracy in representing the epidemiological dynamics. Hence, in this study, we proposed a novel algorithm that utilizes an agent-based simulation (ABS) \cite{software_hard_1,software_hard_2} representation of a norm-based spatio-temporal model to obtain a numerically similar graph-based spatio-temporal model. We evaluate the proposed model on three different temporal epidemiological models as well as three different pandemic settings to explore the algorithm's performance.

The rest of the paper is organized as follows. Section \ref{sec:model} outlines the model definition and ABS representation. Section \ref{sec:algorithm} provides a formal description of the norm-based to graph-based representation conversion task and possible heuristic numerical-based solutions to this task. Section \ref{sec:experiments} describes the experimental setup used to evaluate the proposed algorithms and the obtained experimental results. Finally, Section \ref{sec:discussion} discusses the results, the proposed algorithm's limitations, and provides possible future work. 

\section{Generic Spatio-Temporal Model Formalization}
\label{sec:model}
In this section, we outline the formalization of the temporal component of the epidemiological model based on the SIR model's structure, the norm-based spatial component, the graph-based spatial component formalization, and the ABS representation of the model.

\subsection{Temporal component formalization}
For the temporal component, let us assume a population of timed finite state machines \cite{fsm} agents, \(P\), such that each agent in the population, \(p \in P\), is defined by a tuple \(p := (\xi, \tau)\) where \(\xi \in \Xi\) is the current epidemiological state of the agent out of the set of all possible epidemiological states (\(\Xi\)) and \(\tau\) is a set of variables used to determinate the additional properties of the agent. 

In addition, the temporal component contains a function \(I_s: P \rightarrow P\) which gets the population of agents and returns the same population of agents after altering the state of each agent in the population based on their current state and change over time. For example, let us consider the ordinary differential equation (ODE) representation of the SIR model \cite{first_sir}: 
\begin{equation}
    \frac{dS(t)}{dt} = - \beta S(t)I(t); \frac{dI(t)}{dt} = \beta S(t)I(t) - \gamma I(t); \frac{dR(t)}{dt} = \gamma I(t), 
    \label{eq:sir}
\end{equation}
where \(S(t), I(t)\), and \(R(t)\) are the number of susceptible, infected, and recovered agents in the population, \(\beta\) is the average infection rate, and \(\gamma\) is the average recovery rate. In this example, the second term of the \(\frac{dI(t)}{dt}\) and the third equation is part of the \(I_s\) function while the term \(\beta S(t) I(t)\) is not as the change in the agents' state is not exclusively depended on time. 

Importantly, we allow \(I_s\) to both add and remove agents to the population. This allows the temporal model to include the birth and death of agents into and from the dynamic, respectively. 

\subsection{Norm-based spatial component formalization}
For the norm-based spatial component, one should consider three main elements: the environment (\(E\)), agents' walk dynamics, and infection dynamics. In this formalization, we assume that the information related to the agent's location, movement, and infection dynamics are described as part of the \(\tau\) parameter. 

Intuitively, the environment element, \(E\), outlines the physical location where the pandemic takes place. For example, for small-scale the environment can be a specific building, and for larger-scale, the environment can be a city or even a country. On top of that, the resolution of the environment and its accuracy are also important parts of it. Following the same example, a one-meter resolution of a building is feasible while the same resolution for a city can be computationally infeasible for the average modern computation system. Similarly, in a building or even room setting, one may take into consideration the agents' breathing patterns \cite{b4} as the population size is small while on the city level, the time of year may be more relevant (and computationally feasible) property to consider. Formally, The environment is presented by continuous space, \(L \in \mathbb{R}^k\) (\(k \in \{1, 2, 3\}\)) is the spatial dimension of the environment), such that point, \(l \in L\), is represented by some set of parameters denoted by \(\{\alpha_1, \dots, \alpha_j\}\) where \(j \in \mathbb{N}\) is the number of parameters. Theoretically, an infinite number of locations are presented in the environment. 

The agent's walk dynamics is a function \(I_w: (p, P, E) \rightarrow (p, E)\) that gets a specific individual, the entire population, and the environment and returns the same individual with the relevant alter to the \(\tau\) parameter related to its location and other variables which related to the fact it moved, as well as updating the relevant points in the environment. 

Finally, the agents' interaction dynamics is a function \(I_i: (P, E) \rightarrow (P, E)\) that gets the population and the environment and returns the population with the relevant alter to the \(\xi\) epidemiological parameter as well as to \(\tau\) while also updating the relevant points in the environment. This function is responsible for capturing the infection performed between agents in the population as well as between an agent and the environment itself. 

\subsection{Graph-based spatial component formalization}
The graph-based spatial component is similar to the norm-based spatial component, where the differentiating is present only in the formalization of the environment itself. Namely, the environment, \(E\), is represented by a graph \(G := (V, E)\) where \(V\) are a set of locations and \(E \subset V \times V\) is a set of edges between these locations. Importantly, the size of \(|V|\) is finite. For the agent's walk dynamics (\(I_w\)) we allow individuals to move from a current location (\(v_c \in V\)) to a neighbor location (\(v_n \in V\)) in a single step in time if and only if \(e = (v_c, v_n) \in E\). Finally, for the interaction dynamics, \(I_i\), we assume that at each location (\(v \in V\)) the individuals are well-mixed. 

\subsection{Agent-based simulation representation}
Following the structure proposed by \cite{teddy_review}, an ABS representation of a spatio-temporal model is presented in Algo 1. Initially, a population of agents (\(P\)) and the environment (\(E\)) are generated. Then, in an iterative manner, until a pre-defined stop condition (\(C\)) is met, three operators are taking place - the spontaneous operator (\(I_s\)), the walk operator (\(I_w\)), and the interaction operator (\(I_i\)). These operators alter the agents' state over time. Commonly, the population's epidemiological state (denoted by \(P_\xi\)), defined to be the vector of agents' states, over time is returned for further analysis.

\begin{algorithm}[!ht]
	\caption{Generic agent-based simulation representation of spatio-temporal epidemiological model.} \label{algo:gen}
	\begin{algorithmic}[1]
	    \STATE $\text{\textbf{Input: }} \text{population }  (P), \text{environment } (E)$
	    \STATE $\text{\textbf{Output: }} \text{Population's state over time }  (A)$
        \STATE $t \leftarrow 1$
	    \WHILE{$C(P, t, E)$}
  	        \STATE  $P \Leftarrow I_s(P)$
                \FOR{$p \in P$}
      	        \STATE  $p \Leftarrow I_w(p, P, E)$
                \ENDFOR
  	        \STATE  $(P, E) \Leftarrow I_i(P, E)$
  	        \STATE  $A[t] \leftarrow P_\xi$
                \STATE $t \leftarrow t + 1$
  	    \ENDWHILE
  \STATE return $A$
	\end{algorithmic}
\end{algorithm}

\section{Norm-based to Graph-based Spatial Component Translation}
\label{sec:algorithm}
Following the formalization in Section \ref{sec:model}, one can notice that the translation of the norm-based to graph-based spatial component is trivial without constraints as one can pick a discretization to the continuous environment which is at least as small as the smallest interaction-related distance and generate a mesh graph. However, this solution ignores the motivation of reducing computational time when transforming from norm-based to graph-based spatial component representation while also not necessarily resulting in analytically isomorphic results. To prove this point, let us again consider the SIR model in Eq. (\ref{eq:sir}) but this time we define the infection rate at time \(t\) between two agents in the population (\(i, j\)) to be \(\beta_{i,j} = \frac{1}{|L_i - L_j|}\) where \(L_x \in \mathbb{R}^k\) is the locations of the \(x\)'s agents at time \(t\) \cite{teddy_pandemic_in_the_field}. Moreover, let us assume the agents in the population apply the following walking dynamics: \(\forall j: L(t+1) = \frac{1}{|P|-1} \sum_{i=1, i \neq j}^{|P|} L_i(t) \). In this case, there is no existing fixed discretization size which is at least as small as the smallest interaction-related distance as the latter converges to zero. 

\subsection{Problem definition}
Following this example, the desired graph-based representation should be both computationally appealing and preserve as much as possible the norm-based dynamics. Hence, we formally define the norm-based to graph-based spatial component translation problem as follows:
\begin{equation}
    \min_{|V|} \min |A_n - A_g|
\end{equation}
where \(A_n\) and \(A_g\) are the population's state over time for the norm-based and graph-based spatial components, respectively. From the computational point of view, one should define the graph, \(G = (V, E)\), and the walk dynamics on the graph. 

\subsection{Heuristic numerical-based solution}
Since the problem is challenging and solving it analytically can be computationally expensive (or even infeasible) even for relatively small cases, we propose a heuristic approach to solve the problem. Namely, let us consider a given case with a norm-based spatial component with some walk dynamics. In order to convert it into a graph-based, one would need to find \(G\) as well as transform the walk dynamics to the constraints enforced by \(G\). Below, we describe four methods for the graph search and three for the walk dynamics approximation, resulting in 12 possible models. 

\subsubsection{Graph search}
For the graph search component, we assume each algorithm gets the locations of the agent population at each step in time and returns the spatial graph (\(G\)). 

\paragraph{Quadtree.}
A Quadtree is a hierarchical data structure used for efficient spatial partitioning of an environment. It recursively subdivides a region of dimension \(k\) into \(2^k\) sub-sections, each representing a smaller portion of the environment \cite{qt1}. This tree-like structure is commonly employed in computer graphics, image processing, and geographical information systems for efficient spatial indexing and searching \cite{qt2}.  In our case, one can compute the Quadtree for each step in time given the locations of the agent population. Formally, let us set the root node of the Quadtree to be the entire environment. In a recursive manner, divide the node into \(2^k\) nodes (for a \(k\) dimensional space) such that each node either does not contain any agent, contains a single agent with can not interact with other agents, or all the agents in the node can interact with each other \cite{qt3}. This process results in a single Quadtree. In order to capture the overall graph, \(G\), we calculate the average Quadtree. In order to calculate the average Quadtree from a vector of Quadtrees, one would perform a component-wise average operation on the corresponding nodes of each Quadtree in the vector. Starting from the root, for each level of the Quadtree, average the values or properties of the nodes in the same position across all Quadtrees in the vector. Repeat this process recursively for each level until you reach the leaves of the Quadtree. Finally, each leave node of the average Quadtree is set to be a vertex (\(v\)) of the graph (\(G\)) and two leave nodes with a common parent node share an edge (\(e\)) between them. 

\paragraph{Genetic algorithm (GA).} A GA is an optimization technique inspired by the process of natural selection, where populations of potential solutions evolve over generations \cite{ga_exp_1,ga_exp_2}. It involves the use of genetic operators such as selection, crossover, and mutation to iteratively improve candidate solutions toward achieving optimal or near-optimal outcomes for a given problem \cite{ga_op_1,ga_op_2,ga_op_3}. In order to use the GA approach, one is required to answer seven modeling decisions: 1) How to define a member (also known as \say{gene}) in the population; 2) how to generate the initial population; 3) what is the stop condition for the evolution process; 4) how to define the mutation operator; 5) how to define the cross-over operator; 6) how to define the selection operator; and 7) how to define the loss function of each member (also known as the \say{fitness function}) \cite{markov_biology1,teddy_eco_alexi}. To this end, we represent each member in the GA's population as a vector of location such that each location is the node's location in the environment. An agent is associated with the node that is the closest to it at any given point in time. Next, the GA's population initialization involves generating an initial population with a random number of nodes as well as their locations, denoted by \(\delta_i\) for the \(i_{th}\) agent. Hence, the loss function (i.e., the \textit{fitness} function) is defined to be:
\begin{equation}
    L(m) := \frac{1}{T}\sum_{t=0}^T ( \frac{1}{N}\sum_{i=1}^n \delta_i).
    \label{eq:graph_search_loss}
\end{equation}
In order to obtain a more computationally appealing initial condition, for an initial number of nodes \(x \in \mathbb{N}\), we compute the set cover using a greedy algorithm for the initial state of the agents' population (i.e., \(t=0\)) \cite{ga_initial}. The stop condition for the algorithm is a pre-defined number of iterations (also known as \say{generation}) \cite{ga_intro}. For each iteration, three operators are performed. First, the mutation operator picks a node's index randomly, with a uniform distribution, and changes its location by adding to its current location a random vector sampled from a pre-defined \(n\)-dimensional normal distribution. Second, the cross-over operator accepts two node location vectors of identical size (\(V_1, V_2\)), picks an index, \(j \in [1, \dots, |V_1|] \) in a uniform distribution and returns \(V_1^{new} = V_1[1, \dots, j] \cup V_2[j+1, \dots, |V_2|\) and \(V_2^{new} = V_2[1, \dots, j] \cup V_1[j+1, \dots, |V_2|\) \cite{cross_over}. Finally, for the selection operator, we adopted the \textit{royalty tournament} operator \cite{selection_operator}, which selects the top $\alpha \in [0, 1]$ graphs from the population according to their loss function. Afterward, the population is sampled (with repetitions) according to their loss function values, i.e., with probability \(p_{select}(m) = \frac{L(m)}{\sum_{m' \in P_i} L(m')}\) where \(P_i\) is the population in the \(i_{th}\) iteration.

\paragraph{Time series X-means (TSxM).} Time series K-means (TSkM) is a clustering algorithm tailored for time-dependent data, aiming to identify homogeneous groups or clusters based on temporal patterns \cite{tskm_intro}. It employs the traditional K-means approach but incorporates dynamic time warping or other distance metrics suitable for capturing similarities in time series data. On top of that, it automatically finds the number of clusters, \(k\), using hyperparameter optimization process. To use the TSxM approach, the location of each agent in the population is treated as a time-series vector, with each value representing the location of the agent in the environment in a single step in time. The locations are normalized to the environment's size to ensure comparability. The number of clusters (i.g., the nodes of the graph \(G\), \(x\), is set to \(1\) and increases up to a pre-defined number \(\epsilon \in \mathbb{N}\). The final value of \(x\) is obtained using the elbow point method \cite{elbow}. For each value of \(x\), a Dynamic Time Warping K-means algorithm \cite{xmeans} is used to cluster the time-series data. As the Dynamic Time Warping K-means algorithm is sensitive to its random initial condition, we run it for each simulation for \(a \in \mathbb{N}\) times, taking the configuration with the smallest \(L_1\) metric value between the agent's true location and the cluster's center of mass. 

\paragraph{Recurrent Neural Network (RNN).} A Recurrent Neural Network (RNN) is a type of neural network architecture designed to process sequential data by maintaining a hidden state that captures information from previous time steps and enables the modeling of temporal dependencies in sequences, making it suitable for tasks such as time-series prediction and classification \cite{rnn}. In particular, for our case where the number of agents in the population may change over time, we can not assume a fixed input space's size. As such, we use two layers of Gated Recurrent Unit (GRUs) such that one obtains the agent population as a vector with arbitrary size followed by two fully connected layers to obtain a fixed-sized and with more computationally appealing feature space, while the latter accepts sequences of the processed agent's population locations and also followed by two layers of fully connected layers. As such, the input of the RNN model is the agent population's locations over time and the output is the graph's nodes' locations. For the loss function, we adopted Eq. (\ref{eq:graph_search_loss}) and for the optimization algorithm, we adopted ADAM \cite{adam_optimizator}.

\subsubsection{Walk dynamics approximation}
For the walk dynamics approximation component, we assume each algorithm gets the locations of the agent population at each step in time and the spatial graph (\(G\)) and returns a walk dynamics function (\(I_w\)). 

\paragraph{Markov chain (MC).}
A Markov chain is a stochastic process where the probability of transitioning to a particular state depends only on the current state, disregarding the sequence of events leading to that state. Intuitively, it is characterized by a set of states and transition probabilities between those states \cite{markov_book,markov_astronomy}. In our case, for each node in the graph (\(v \in V\)) and agent (\(p \in P\)) that is currently located in the node, let us assume the MC's state is represented by the agent's state, the environment's state, and the epidemiological's state distribution of the agents in each node \(v_n \in V\) such that \(e = (v, v_n) \in E\). Notably, this approach ignores any additional properties the agents might have in favor of the common (epidemiological) ground.

\paragraph{Multi-agent classification with AutoEncoder (MAC-AE).} The multi-agent classification approach stands for the idea that a classification task can be solved by multiple models, each one governing a sub-set of the task and they are autonomous concerning each other. In our case, each node of the graph (\(v \in V\)) is associated with an ML classification model. Each such ML model obtains the state of an agent currently located in the node, the environment, and an entire population. However, as the size of the entire population can alter over time, a fully connected attention AutoEncoder model \cite{autoencoder,teddy_cancer} is used to obtain a fixed-length representation of the population data. As such, the model is trained on tuples of the agent's state, environment, and encoded population representation while the target column is the agent's location, as represented by the graph node's index, in the next step in time. In order to find the best ML model for each node, we used an automatic machine learning (AutoML) approach. Namely, we adopted the Tree-Based Pipeline Optimization Tool (TPOT), a genetic algorithm-based automatic ML library \cite{tpot}. TPOT produces a full ML pipeline, including feature selection engineering, model selection, model ensemble, and hyperparameter tuning which is close to optimal for relatively long enough computational time \cite{teddy_bio_inspired_auto_ml}.

\paragraph{Deep reinforcement learning (DRL).} Deep Reinforcement Learning (DRL) is a data-driven algorithm that combines reinforcement learning with deep neural networks to enable agents to learn complex behaviors and decision-making processes \cite{drl_teddy,drl_1,drl_2,syntheye}. It involves training artificial agents to interact with an environment, receive feedback in the form of rewards or penalties, and leverage deep neural networks to approximate optimal strategies for maximizing cumulative rewards over time \cite{rl_review}. For our case, the walk dynamics are represented in the same manner as the MAC-AE method. However, rather than using an AutoML model with AE for each node of the graph (\(v \in V\)), we use a single DRL model for the entire graph (\(G\)). To be exact, we adopt the Proximal Policy Optimization (PPO) \cite{ppo} which considers one of the state-of-the-art DRL algorithms for a wide range of tasks. Specifically, we use a fully connected neural network at the base of the PPO algorithm.

\section{Results}
\label{sec:results}
Based on the algorithms presented in Section \ref{sec:algorithm}, an experimental setup is defined for both real-world and synthetic cases such that the real-world cases are used to evaluate the algorithms' performances in realistic scenarios while the synthetic cases are used for further sensitivity analysis of different spatial conditions.  

\subsection{Experimental Setup}
\label{sec:experiments}
In order to evaluate the proposed algorithms, we adopted three temporal component dynamics as well as three norm-based spatial components with increasing levels of complexity for both. For the temporal component, we used SIR \cite{first_sir}, SIRD with two age groups \cite{teddy_pandemic_management}, and two-strain SIR models \cite{teddy_multi_strain}. Simply put, SIR is considered the most simplistic model with only three states - susceptible (S), infected (I), and recovered (R) individuals where susceptible individuals become infected from interaction with infected individuals while infected slowly recover over time until they fully recovered and transform to the recovered epidemiological state where they immune to re-infection. The SIRD with two age groups introduces a dead state (denoted by \(D\)) where individuals who are infected have some probability of dying rather than recovering. Moreover, it extends the SIR model by dividing the population into two age groups (for example, children and adults) with asymmetric infection rates and unique recovery durations. Finally, the two-strain SIR assumes two pathogens are causing a pandemic in parallel and individuals can be infected and/or recovered from either or both of them at any given moment. A formal ODE and ABS descriptions of each temporal model are provided in the Appendix. 

Since each epidemiological scenario is known to have unique parameter values, we set a biologically relevant range for the synthetic cases while adopting the values presented in the original works of the real-world cases, following the values for COVID-19 \cite{review_covid}. Table \ref{table:param_vals} summarizes the parameters and their values for all four cases. 

\begin{table}[!ht]
\centering
\begin{tabular}{cccccc}
\hline \hline
\textbf{Symbol} & \textbf{Parameter} & \textbf{Synthetic} & \textbf{Airport} & \textbf{Resturant} & \textbf{Bus} \\ \hline \hline
 \(\beta\) & Infection rate in a hour [\(t\)] & \(0.037 - 0.1 \) & \(0.05 - 0.1 \) & \(0.062 - 0.079 \) & \(0.048 - 0.069 \)  \\
 \(\gamma\) & Recovery duration in a hour [\(t\)] & \(120 - 240 \) & \(144 - 170 \) & \(144 - 170 \) & \(144 - 170\)  \\
 \(\rho\) & Recovery rate [\(1\)] & \(0.9 - 0.99 \) & \(0.98 - 1\) &\(0.98 - 1\) & \(0.98 - 1\) \\
 \(\psi\) & Symptomatic rate [\(1\)] & \(0.05 - 0.20 \) & \(0.1 - 0.15 \) & \(0.1 - 0.15 \) & \(0.1 - 0.15 \)  \\ \hline \hline
\end{tabular}
\caption{A summary of the epidemiological model's parameters' values for the synthetic and three real-world cases. }
\label{table:param_vals}
\end{table}

Moreover, the proposed algorithms require that some hyperparameter would be configured before being used. These hyperparameters are closely related to the algorithm's performance and commonly different for different tasks to obtain the algorithm's optimal performance \cite{hyperparamters_are_hard}. As such, we used the default suggested hyperparameters' values from the code libraries we adopted in Python. 

\subsubsection{Real-world cases}
\begin{figure}[!ht]
    \centering
    \includegraphics[width=0.99\textwidth]{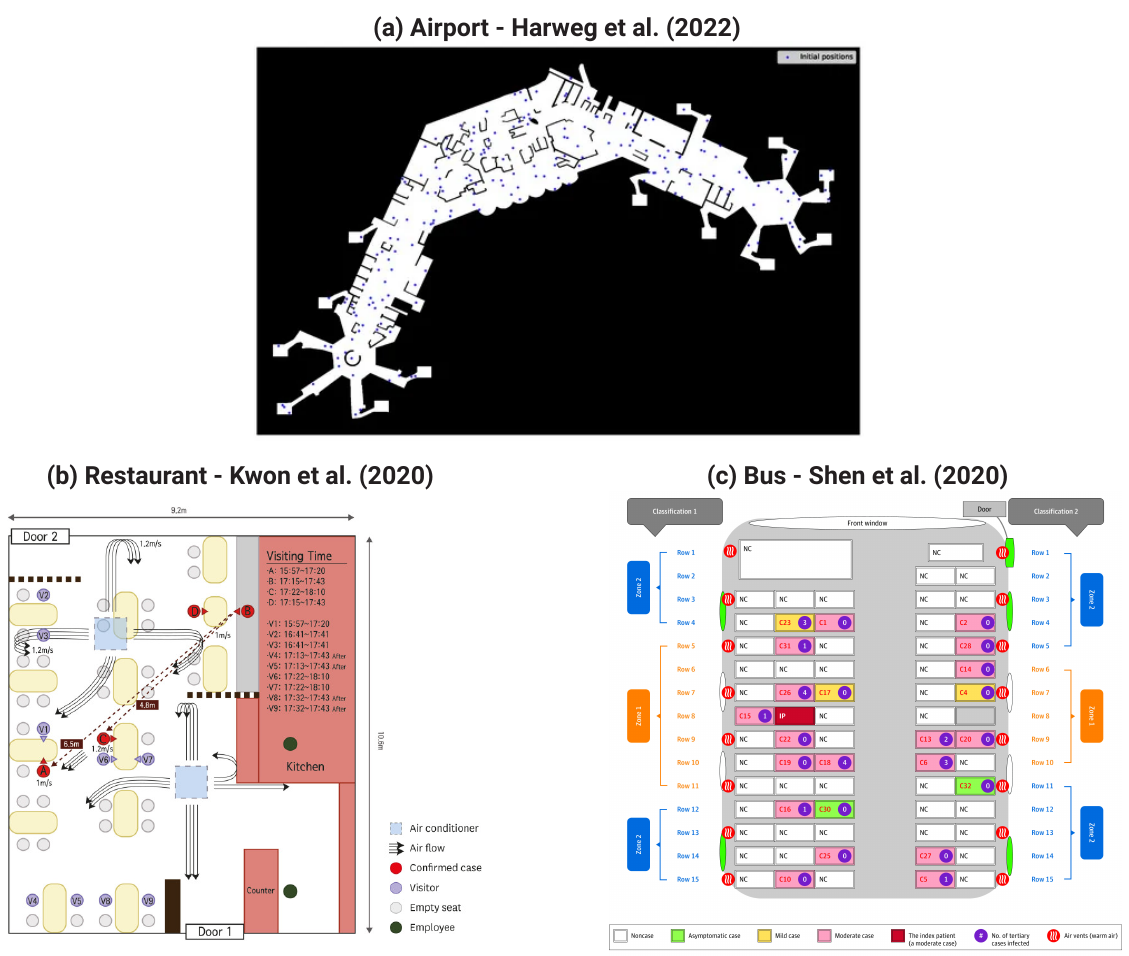}
    \caption{A schematic view of the locations used for the experiments. Images taken from \cite{airport_case,bus_case,restaurant_case}.}
    \label{fig:expertiment_setup}
\end{figure}

For the real-world cases, we adopted three locations and walks: airport \cite{airport_case}, restaurant \cite{restaurant_case}, and bus \cite{bus_case}. Fig. \ref{fig:expertiment_setup} shows schematic views of the three locations as presented by \cite{airport_case,bus_case,restaurant_case}. In order to make the representation of all three locations identical, we used pixel-level measurements of the provided images alongside the scale declared by the authors, generating a grid of (a) 2,5 meters, (b) 0.5 meters, and (c) 0.5 meters. Despite the fact that height plays a role in the infection dynamics \cite{b4}, it is shown only when taking airflow dynamics into account, which we do not do in this experiment. Hence, we ignore any height-related information that may be in these settings.

In addition, we adopted the walk dynamics presented in the original papers. Simply put, each agent has a unique set of locations in the space which is influenced by the simulation's step in time as well as the presence of other agents. In the airport, agents randomly obtain a path from the entrance and a gate and move at a constant speed until they arrive at the gate. For the restaurant, some agents were allocated to sit, at random, according to the time of the day in the simulation while small, pre-defined, agents performed random walks. In a similar manner, agents in the bus case enter from the top door, picking a free sit at random, walking to the seat, waiting for a random number of simulation steps, moving to the exit door, and existing. 

Importantly, as not the entire population is present in the dynamics at any point in time, as required by the SIR-based temporal component, we assume individuals getting and leaving the scene are not interacting with each other and only can recover (or die) if they are infected. We pick the individuals to enter the scene at random, in a uniformly distributed manner. 

\subsubsection{Synthetic cases}
\begin{figure}
    \centering
    \includegraphics[width=0.99\textwidth]{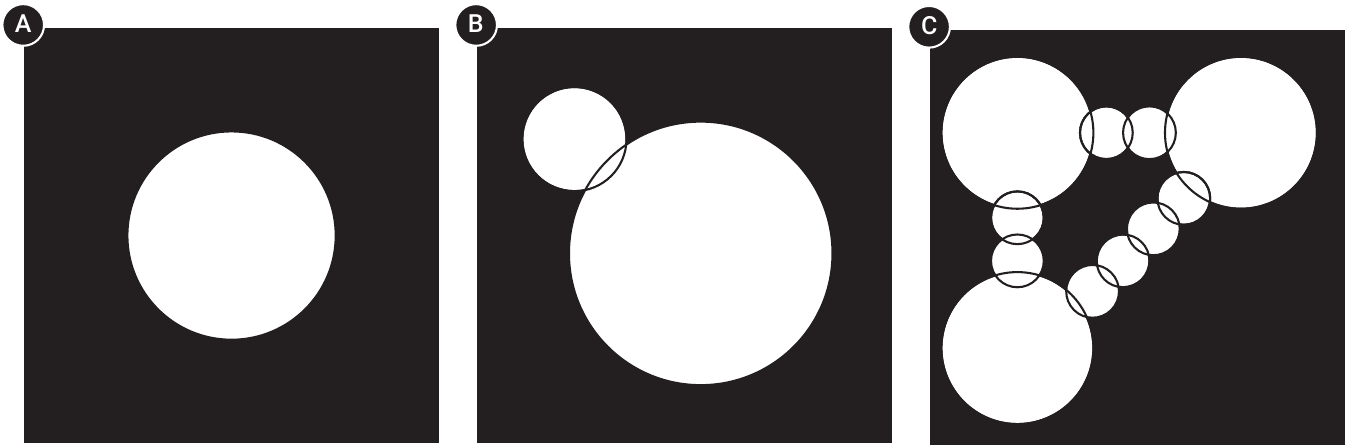}
    \caption{Examples for (a) the simplest configuration; (b) configuration with different sub-locations with different densities; and (c) spatially-enforced walk patterns using the synthetic spatial component formalization.}
    \label{fig:synt_spatio_example}
\end{figure}

In order to explore the proposed algorithms' sensitivity to multiple spatial components of the space, we decided to keep the spatial configuration simple and highly adaptable. To this end, we allocate an on a plane, a set of non-pairwise distinct circles. Each circle is defined by its radius and center location. As such, the spatial configuration of each synthetic case is defined by a set of circles and their properties. This allows the generation of a wide space of configurations while keeping the seed relatively simple. Moreover, it allows for control of the average density of the population, enforces walking patterns, etc. An example of the simplest configuration, a configuration with different sub-location with different densities, and spatially-enforced walk patterns are shown in panels a, b, and c, respectively, in Fig. \ref{fig:synt_spatio_example}. For simplicity, in these examples, we assume individuals perform random walks and can occupy the same location. To capture complex enough settings, we set the number of circles to be between 1 and 40 and the radius to be between 2 and 25 meters. 

We assume individuals follow random walks, but have a \say{pulling} power towards their original spawn location which is reversibly proportional to their distance up to some distance \(\delta\).  

\subsection{Performance analysis}
We start by exploring the performance of all twelve possible combinations of the graph search and the walk dynamics approximation for the synthetic cases. As such, we generate \(n=10000\) random synthetic cases where the population size is set to 1000 individuals.   Table \ref{table:performance_synetetic} summarizes the results of this analysis such that the performance represents the average agreement between the norm-based population's state and the graph-based population's state over time. Noteworthy, the relatively simplistic Quadtree results in worse performance compared to the other methods. The TSxM archives slightly higher but comparable performance, on average, compared to the GA and RNN graph search algorithms. For the walk dynamics approximation, the MC algorithm, on average, archives worse performance compared to the MAC-AE and DRL algorithms which achieve similar performances. Generally speaking, a large number of nodes and edges, on average, corresponds to higher performance, as one could expect.  

\begin{table}[!ht]
\centering
\begin{tabular}{ccccc}
\hline \hline
\textbf{Graph search} & \textbf{Walk dynamics approximation} & \textbf{Performance} & \textbf{\(|V|\)} & \textbf{\(|E|\)} \\ \hline \hline
\multirow{3}{*}{Quadtree} & MC & \(0.73 \pm 0.06 \) & \(72.14 \pm 39.14 \) & \(5.14 \pm 3.88\)  \\
 & MAC-AE & \(0.76 \pm 0.05 \) & \(65.38 \pm 35.15 \) & \(5.48 \pm 3.26 \) \\
 & DRL & \(0.68 \pm 0.10 \) & \(46.03 \pm 51.09 \) & \(5.95 \pm 4.78 \) \\ \hline
\multirow{3}{*}{GA} & MC & \(0.84 \pm 0.07\) & \( 75.19 \pm 37.13 \)  & \(6.02 \pm 3.70\)  \\
 & MAC-AE & \(0.88 \pm 0.08\) & \(62.29 \pm 42.08\) & \(4.89 \pm 2.90\) \\
 & DRL & \(0.82 \pm 0.07\) & \(67.58 \pm 40.37\) & \(5.12 \pm 3.23\) \\ \hline
\multirow{3}{*}{TSxM} & MC & \(0.84 \pm 0.07\) & \(68.14 \pm 40.05\) & \(4.19 \pm 3.05\)  \\
 & MAC-AE & \(0.90 \pm 0.09\) & \(77.08 \pm 34.40\) & \(4.65 \pm 3.15\) \\
 & DRL & \(0.89 \pm 0.10\) & \(75.19 \pm 37.11\) &  \(4.95 \pm 3.80\) \\ \hline
\multirow{3}{*}{RNN} & MC & \(0.83 \pm 0.05\) & \(81.06 \pm 43.72\) & \(4.22 \pm 2.96\)  \\
 & MAC-AE & \(0.85 \pm 0.06\) & \(83.71 \pm 41.18\) & \(4.08 \pm 3.53\)  \\
 & DRL & \(0.86 \pm 0.05\) & \(68.26 \pm 40.01\) & \(4.21 \pm 3.04\) \\ \hline \hline
\end{tabular}
\caption{Performance analysis for the synthetic cases divided into the 12 graph search and walk dynamics approximation options. All models trained for \(n_{train} = 900\) random cases and evaluated on \(n_{test} = 100\) random cases such that the performance, \(|V|\), and \(|E|\) are shown as the mean \(\pm\) standard deviation of the test set. }
\label{table:performance_synetetic}
\end{table}

Table \ref{table:performance_real_world} presents the performance of the best graph search and best walk dynamics approximation. Intriguingly, there is no single graph search or waly dynamics approximation algorithm that outperforms all others for all three cases. Nonetheless, the combination of TSxM with the MAC-AE algorithms seems to have good performance compared to the remaining options. In all three cases, the performance achieved is higher than 0.9, on average. Moreover, the larger geometrical configuration, the airport case, required an order of magnitude more nodes compared to the relatively small geometrical configurations, the restaurant and bus. Similarly, the average number of edges (4.32) is also larger for the airport case where individuals have a more chaotic behavior compared to the bus (3.77) and restaurant (2.89).  

\begin{table}[!ht]
\centering
\begin{tabular}{cccccc}
\hline \hline
\textbf{Case} & \textbf{Best graph search} & \textbf{Best walk dynamics approximation} & \textbf{Performance} & \(|V|\) & \(|E|\) \\ \hline \hline
Airport & GA & MAC-AE & \(0.93 \pm 0.05 \) & \(173.28 \pm 11.35 \) & \(4.32 \pm 1.48 \) \\
Restaurant & TSxM & MAC-AE & \(0.95 \pm 0.04 \) & \(18.13 \pm 2.04 \) &  \(2.89 \pm 0.66 \) \\
Bus & TSxM & DRL & \(0.92 \pm 0.02 \)  & \(26.41 \pm 3.16 \) &  \(3.77 \pm 0.34 \) \\ \hline \hline
\end{tabular}
\caption{Performance analysis for real-world cases. All 12 model combinations are trained on \(n_{train} = 900\) random initial conditions of each real-world case and evaluated on \(n_{test} = 100\) other random initial conditions such that the performance, \(|V|\), and \(|E|\) are shown as the mean \(\pm\) standard deviation of the test set for the best combination of graph search and walk dynamics approximation. }
\label{table:performance_real_world}
\end{table}

\subsection{Sensitivity analysis}
While the results do not reveal a clear winner for all cases, the TSxM with MAC-AE shows superior performance compared to the other algorithms, on average. As such, we explore its performance sensitivity to several temporal and spatial properties, as follows. First, we assume a circle with a radius of 50 meters spatial configuration and alter the density by changing the population size from \(3.14 \cdot 10^2\) to \(3.14 \cdot 10^4\). Second, we assume the radius of all circles is constant at equal to 2 meters while the number of circles defining the geometrical configuring ranges from 10 and 100. Third, we explore the influence of different levels of infection ranging the infection rate (\(\beta\)) from \(0.037\) to \(0.37\). Fourth, we explore the influence of different recovery durations (\(\gamma\)), ranging from \(60\) to \(380\) hours. Table \ref{table:sens} summarizes the obtained results for \(n=1000\) random cases, picked in a uniformly distributed manner. Notably, the sensitivity analysis of the spatially-related features (the first two rows) resulted in lower performance compared to the temporally-related features (the last two rows). Moreover, their standard deviation as also higher (0.02 and 0.03 vs. 0.01 and 0.01, respectively). These indicating the proposed method is less sensitive to temporal changes compared to spatial ones.

\begin{table}[!ht]
\centering
\begin{tabular}{lcc}
\hline \hline
\textbf{Parameter} & \textbf{Range} & \textbf{Performance} \\ \hline\hline
Population size (\(N\)) & 314-31400 & \(0.95 \pm 0.02\) \\
Number of circles defining the geometrical configuration & 10-100 & \(0.94 \pm 0.03\) \\
Infection rate (\(\beta\)) & 0.037-0.37 & \(0.97 \pm 0.01\) \\
Recovery duration in hours (\(\gamma\)) & 60-380 & \(0.97 \pm 0.01 \) \\ \hline\hline
\end{tabular}
\caption{Sensitivity analysis of the TSxM with MAC-AE to spatial and temporal changes. The results are shown as the mean \(\pm\) standard deviation of \(n=1000\) random cases. }
\label{table:sens}
\end{table}

\section{Discussion}
\label{sec:discussion}
In this study, we proposed a novel approach for translating norm-based spatial components to graph-based representations in the context of epidemiological modeling. Our approach balances computational efficiency with the preservation of norm-based dynamics, thereby facilitating more efficient simulations while maintaining analytical fidelity. 

Our experimental results demonstrated the effectiveness of various algorithms for translating norm-based spatial components to graph-based representations. Through extensive experimentation with synthetic and real-world scenarios, we evaluated twelve combinations of graph search and walk dynamics approximation algorithms. The analysis revealed promising performance across different scenarios. Notably, the Time Series X-means (TSxM) algorithm combined with Multi-agent Classification with AutoEncoder (MAC-AE) for walk dynamics approximation consistently demonstrated robust performance across both synthetic and real-world cases.

Practically, the proposed approach has significant implications for public health policy and epidemic management. By enabling more efficient simulations of epidemic spread in real-world environments such as airports, restaurants, and buses, our methodology can inform decision-making processes aimed at mitigating disease transmission \cite{teddy_chaos,end_1}. Moreover, the ability to accurately model spatial dynamics in diverse settings enhances our understanding of epidemic dynamics and facilitates the development of targeted intervention strategies \cite{end_2,end_3}.

The study's proposed methodology introduces promising advancements but comes with several limitations. Mainly, the proposed method is not validated against real-world data, leaving uncertainties about the accuracy and reliability of the proposed approach in predicting actual pandemic dynamics. However, this limitation is in the field of epidemiological modelling as the required data is usually scares or even unavailable. Second, the proposed method is not interpretable nor explainable as it solves a multi-dimensional and complex task using a heuristic or ML approach, making the proposed approach a \textit{black-box}. This fact may limit the ability of policymakers to put their faith in this solution by itself or as part of a larger epidemiological model \cite{end_4,end_5}. Moreover, the study predominantly assumes a static environment throughout the spatio-temporal simulations, disregarding potential changes in the spatial landscape that could influence epidemic dynamics. As such, future research should aim to refine the proposed solutions and address these limitations for a more comprehensive understanding of their practical implications.

\section*{Declarations}
\subsection*{Funding}
This research received no specific grant from any funding agency in the public, commercial, or not-for-profit sectors.

\subsection*{Conflicts of interest/Competing interests}
The author does not have a conflict of interest to declare.

\subsection*{Acknowledgment}
The author wishes to thank Sveta Hardak-Nissan for teaching me the professional maturity required for fundamental research.
 
\bibliography{biblio}

\begin{thebibliography}{10}

\bibitem{pandemic_important}
Andrea~Alberto Conti.
\newblock {Historical and methodological highlights of quarantine measures: from ancient plague epidemics to current coronavirus disease (COVID-19) pandemic}.
\newblock {\em Acta bio-medica : Atenei Parmensis}, 91(2):226--229, 2020.

\bibitem{pandemic_duration}
A.~Brodeur, D.~Gray, A.~Islam, and S.~Bhuiyan.
\newblock A literature review of the economics of covid-19.
\newblock {\em IZA Discussion Paper No. 13411, Available at SSRN: https://ssrn.com/abstract=3636640}, 2020.

\bibitem{who_problem}
Eurosurveillance~Editorial Team.
\newblock Note from the editors: World health organization declares novel coronavirus (2019-ncov) sixth public health emergency of international concern.
\newblock {\em Euro Surveill}, 25:200131e, 2020.

\bibitem{intro_pandemic_example_1}
T.~K. Mackey and B.~A. Liang.
\newblock Lessons from sars and h1n1/a: employing a who–wto forum to promote optimal economic-public health pandemic response.
\newblock {\em Journal of Public Health Policy}, 33:119--139, 2012.

\bibitem{teddy_labib_financing_pandemic}
L.~Shami and T.~Lazebnik.
\newblock Financing and managing epidemiological-economic crises: Are we ready for another outbreak?
\newblock {\em Journal of Policy Modeling}, 45(1):74--89, 2023.

\bibitem{intro_pandemic_example_2}
R.~J. De~Groot, S.~C. Baker, R.~S. Baric, C.~S. Brown, C.~Drosten, L.~Enjuanes, R.~A.~M. Fouchier, M.~Galiano, A.~E. Gorbalenya, Z.~A. Memish, S.~Perlman, L.~L.~M. Poon, E.~J. Snijder, G.~M. Stephens, P.~C.~Y. Woo, A.~M. Zaki, M.~Zambon, and J.~Ziebuhr.
\newblock Middle east respiratory syndrome coronavirus (mers-cov): announcement of the coronavirus study group.
\newblock {\em Journal of Virology}, 87:7790–7792, 2013.

\bibitem{example_influenza}
A~D. Iuliano and et~al.
\newblock Estimates of global seasonal influenza-associated respiratory mortality: a modelling study.
\newblock {\em The Lancet}, 391(10127):1285--1300, 2018.

\bibitem{social_distance_1}
C.~Jones, T.~Philippon, and V.~Venkateswaran.
\newblock Optimal mitigation policies in a pandemic: Social distancing and working from home.
\newblock {\em The Review of Financial Studies}, 34(11):5188--5223, 2021.

\bibitem{sir_clinical}
D.~Acemoglu, V.~Chernozhukov, I.~Werning, and M.~D. Whinston.
\newblock Optimal targeted lockdowns in a multigroup sir model.
\newblock {\em American Economic Review: Insights}, 3(4):487--502, 2021.

\bibitem{review_zika}
A.~Wiratsudakul, P.~Suparit, and C.~Modchang.
\newblock Dynamics of zika virus outbreaks: an overview of mathematical modeling approaches.
\newblock {\em PeerJ}, 2018.

\bibitem{review_covid}
A.~Adiga, D.~Dubhashi, B.~Lewis, M.~Marathe, S.~Venkatramanan, and A.~Vollikanti.
\newblock Mathematical models for covid‑19 pandemic: A comparative analysis.
\newblock {\em J. Indian Inst. Sci.}, 100(4):793--807, 2020.

\bibitem{stat_model_1}
A.~Desai, M.~Kraemer, S.~Bhatia, A.~Cori, P.~Nouvellet, M.~Herringer, E.~Cohn, M.~Carrion, J.~Brownstein, L.~Madoff, and B.~Lassmann.
\newblock Real-time epidemic forecasting: challenges and opportunities.
\newblock {\em Health Security}, 17(8):268--275, 2019.

\bibitem{different_approach_from_sir}
B.~Ivorra, M.~R. Ferrandez, M.~Vela-Perez, and A.~M. Ramos.
\newblock Mathematical modeling of the spread of the coronavirus disease 2019 (covid-19) taking into account the undetected infections. the case of china.
\newblock {\em Commun Nonlinear Sci Numer Simulat}, 2020.

\bibitem{different_approach_from_sir_2}
J.~B. Long and J.~M. Ehrenfeld.
\newblock The role of augmented intelligence (ai) in detecting and preventing the spread of novel coronavirus.
\newblock {\em Journal of Medical Systems}, 44, 2020.

\bibitem{stat_model_6}
R.~Salgotra, M.~Gandomi, and A.~H. Gandomi.
\newblock Time series analysis and forecast of the covid-19 pandemic in india using genetic programming.
\newblock {\em Chaos Solitons Fractals}, 138:109945, 2020.

\bibitem{stat_model_5}
P.~Agarwal and K.~Jhajharia.
\newblock Data analysis and modeling of covid-19.
\newblock {\em Journal of Statistics and Management Systems}, 24(1):1, 2021.

\bibitem{models_good_1}
A.~R. Tuite, D.~N. Fisman, and A.~L. Greer.
\newblock Mathematical modelling of covid-19 transmission and mitigation strategies in the population of ontario, canada.
\newblock {\em CMAJ}, 192:E497--E505, 2020.

\bibitem{models_good_2}
J.~C. Miller.
\newblock Mathematical models of sir disease spread with combined non-sexual and sexual transmission routes.
\newblock {\em Infectious Disease Modelling}, 2:35--55, 2017.

\bibitem{models_good_3}
T.~Berge, J.M.-S. Lubuma, G.M. Moremedi, N.~Morris, and R.~Kondera-Shava.
\newblock A simple mathematical model for ebola in africa.
\newblock {\em Journal of Biological Dynamics}, 11(1):42--74, 2017.

\bibitem{miller2017mathematical}
Joel~C Miller.
\newblock Mathematical models of {SIR} disease spread with combined non-sexual and sexual transmission routes.
\newblock {\em Infectious Disease Modelling}, 2(1):35--55, 2017.

\bibitem{sir_popular_5}
M.~Al-Raeei.
\newblock The forecasting of covid-19 with mortality using {SIRD} epidemic model for the united states, russia, china, and the syrian arab republic.
\newblock {\em AUO Advances}, 10(6), 2020.

\bibitem{sir_popular_4}
J.~Fernández-Villaverde and C.~I. Jones.
\newblock Estimating and simulating a {SIRD} model of {COVID}-19 for many countries, states, and cities.
\newblock Working Paper 27128, National Bureau of Economic Research, 2020.

\bibitem{infection_graph_main}
G.~Ellison.
\newblock Implications of heterogeneous {SIR} models for analyses of {COVID}-19.
\newblock {\em National Bureau of Economic Research}, page Working paper 27373, 2020.

\bibitem{first_sir}
W.~O. Kermack and A.~G. McKendrick.
\newblock A contribution to the mathematical theory of epidemics.
\newblock {\em Proceedings of the Royal Society}, 115:700–721, 1927.

\bibitem{optimization_method_1}
R.~H. Byrd, P.~Lu, J.~Nocedal, and C.~Zhu.
\newblock A limited memory algorithm for bound constrained optimization.
\newblock {\em SIAM Journal on scientific computing}, 16(5):1190--1208, 1995.

\bibitem{optimization_method_1b}
J.~Friedman, P.~Liu, C.~E. Troeger, A.~Carter, R.~C. Reiner, R.~M. Barber, J.~Collins, S.~S. Lim, D.~M. Pigott, T.~Vos, S.~I. Hay, C.~J.~L. Murray, and E.~Gakidou.
\newblock Predictive performance of international covid-19 mortality forecasting models.
\newblock {\em Nature Communications}, 12:2609, 2021.

\bibitem{teddy_economic}
T.~Lazebnik, L.~Shami, and S.~Bunimovich-Mendrazitsky.
\newblock Spatio-temporal influence of non-pharmaceutical interventions policies on pandemic dynamics and the economy: The case of covid-19.
\newblock {\em Research Economics}, 2021.

\bibitem{spatio_1}
R.~Baber.
\newblock Pandemics: learning from the past.
\newblock {\em Climacteric}, 23(3):211--212, 2020.

\bibitem{spatio_2}
M.~Andraud, N.~Hens, C.~Marais, and P.~Beutels.
\newblock Dynamic epidemiological models for dengue transmission: A systematic review of structural approaches.
\newblock {\em Plos One}, 7(11):1--14, 11 2012.

\bibitem{spatio_3}
A.~A. Conti.
\newblock {Historical and methodological highlights of quarantine measures: from ancient plague epidemics to current coronavirus disease (COVID-19) pandemic}.
\newblock {\em Acta bio-medica : Atenei Parmensis}, 91(2):226--229, 2020.

\bibitem{b1}
Z.~Peng, P.~A.~L. Rogas, E.~Kropff, W.~Bahnfleth, G.~Buonanno, S.~J. Dancer, J.~Kurnitski, Y.~Li, M.~G. L.~C. Loomans, L.~C. Marr, L.~Morawska, W.~Nazaroff, C.~Noakes, X.~Querol, C.~Sekhar, R.~Tellier, T.~Greenhalgh, L.~Bourouiba, A.~Boerstra, J.~W. Tang, S.~L. Miller, and J.~L. Jimenez.
\newblock Practical indicators for risk of airborne transmission in shared indoor environments and their application to covid-19 outbreaks.
\newblock {\em Environmental Science and Technology}, 56:1125--1137, 2020.

\bibitem{b5}
T.~Lazebnik and A.~Alexi.
\newblock Comparison of pandemic intervention policies in several building types using heterogeneous population model.
\newblock {\em Communications in Nonlinear Science and Numerical Simulation}, 107(4):106176, 2022.

\bibitem{b2}
T.~Fukuoka and K.~Ito.
\newblock Exposure risk assessment by coupled analysis of cfd and sir model in enclosed space.
\newblock {\em AIVC}, 2010.

\bibitem{b3}
Y.~Chen, X.~Liang, T.~Hong, and X.~Luo.
\newblock Simulation and visualization of energy-related occupant behavior in office buildings.
\newblock {\em Building Simulation}, 10:785–798, 2017.

\bibitem{spatial_graph_intro}
N.~Masuda and P.~Holme.
\newblock Temporal network epidemiology.
\newblock {\em Springer: Singapore}, 2017.

\bibitem{spatial_graph_1}
P.~Holme.
\newblock Fast and principled simulations of the sir model on temporal networks.
\newblock {\em Plos One}, 16(2):e0246961, 2021.

\bibitem{spatial_graph_1a}
I.~Z. Kiss, J.~C. Miller, and P.~L. Simon.
\newblock Mathematics of epidemics on networks.
\newblock {\em Cham: Springer}, 2017.

\bibitem{alexi}
A.~Alexi, A.~Rosenfeld, and T.~Lazebnik.
\newblock A security games inspired approach for distributed control of pandemic spread.
\newblock {\em Advanced Theory and Simulations}, 6(2):2200631, 2023.

\bibitem{norm_based_1}
A.~Viguerie, G.~Lorenzo, F.~Auricchio, D.~Baroli, T.~J.~R. Hughes, A.~Patton, A.~Reali, T.~E. Yankeelov, and A.~Veneziani.
\newblock Simulating the spread of covid-19 via a spatially-resolved susceptible–exposed–infected–recovered–deceased (seird) model with heterogeneous diffusion.
\newblock {\em Applied Mathematics Letters}, 111:106617, 2021.

\bibitem{norm_based_2}
M.~Bognanni, H.~Doug, D.~Kolliner, and K.~Mitman.
\newblock Economics and epidemics: Evidence from an estimated spatial econ-sir model.
\newblock {\em Finance and Economics Discussion Series 2020-091. Washington: Board of Governors of the Federal Reserve System}, 2020.

\bibitem{numerical_software_compare}
R.~Connell, P.~Dawson, and S.~Alex.
\newblock Comparison of an agent-based model of disease propagation with the generalised sir epidemic model.
\newblock page ADA510899, 2009.

\bibitem{software_hard_1}
D.~Chumachenko, V.~Dobriak, M.~Mazorchuk, I.~Meniailov, and K.~Bazilevych.
\newblock On agent-based approach to influenza and acute respiratory virus infection simulation.
\newblock In {\em 2018 14th International Conference on Advanced Trends in Radioelecrtronics, Telecommunications and Computer Engineering (TCSET)}, pages 192--195, 2018.

\bibitem{software_hard_2}
J.~D. Priest, A.~Kishore, L.~Machi, C.~J. Kuhlman, D.~Machi, and S.~S. Ravi.
\newblock Csonnet: An agent-based modeling software system for discrete time simulation.
\newblock In {\em 2021 Winter Simulation Conference (WSC)}, pages 1--12, 2021.

\bibitem{fsm}
V.~S. Alagar and K.~Periyasamy.
\newblock {\em Extended Finite State Machine}, pages 105--128.
\newblock Springer London, 2011.

\bibitem{b4}
T.~Lazebnik and A.~Alexi.
\newblock High resolution spatio-temporal model for room-level airborne pandemic spread.
\newblock {\em Mathematics}, 11(2):426, 2023.

\bibitem{teddy_review}
T.~Lazebnik.
\newblock Computational applications of extended sir models: A review focused on airborne pandemics.
\newblock {\em Ecological Modelling}, 483:110422, 2023.

\bibitem{teddy_pandemic_in_the_field}
T.~Lazebnik.
\newblock Cost-optimal seeding strategy during a botanical pandemic in domesticated fields.
\newblock {\em arXiv}, 2023.

\bibitem{qt1}
C.H. Chien and J.~K. Aggarwal.
\newblock Identification of 3d objects from multiple silhouettes using quadtrees/octrees.
\newblock {\em Computer Vision, Graphics, and Image Processing}, 36(2):256--273, 1986.

\bibitem{qt2}
H.~Samet.
\newblock The quadtree and related hierarchical data structures.
\newblock {\em ACM Computing Surveys}, 16(2):187--260, 1984.

\bibitem{qt3}
A.~Krekov, J.~Gruninger, R.~Schlonvoigt, and J.~Kruger.
\newblock Towards in situ visualization of extreme-scale, agent-based, worldwide disease-spreading simulations.
\newblock {\em SIGGRAPH Asia 2015 Visualization in High Performance Computing}, 7:1--4, 2015.

\bibitem{ga_exp_1}
L.~Bo and L.~Rein.
\newblock Comparison of the luus–jaakola optimization procedure and the genetic algorithm.
\newblock {\em Engineering Optimization}, 37(4):381--396, 2005.

\bibitem{ga_exp_2}
A.~Ghaheri, S.~Shoar, M.~Naderan, and S.~S. Hoseini.
\newblock The applications of genetic algorithms in medicine.
\newblock {\em Oman Med J.}, 30(6):406--416, 2005.

\bibitem{ga_op_1}
L.~Davis.
\newblock Applying adaptive algorithms to epistatic domains.
\newblock {\em Proceedings of the international joint conference on artificial intelligence}, pages 162--164, 1985.

\bibitem{ga_op_2}
Z.~W. Bo, L.~Z. Hua, and Z.~G. Yu.
\newblock Optimization of process route by genetic algorithms.
\newblock {\em Robotics and Computer-Integrated Manufacturing}, 22:180--188, 2006.

\bibitem{ga_op_3}
A.~B.~A. Hassanat and E.~Alkafaween.
\newblock On enhancing genetic algorithms using new crossovers.
\newblock {\em International Journal of Computer Applications in Technology}, 55(3), 2017.

\bibitem{markov_biology1}
A.~E. Nix and M.~D. Vose.
\newblock Modeling genetic algorithms with {M}arkov chains.
\newblock {\em Annals of Mathematics and Artificial Intelligence}, 5(1), 1992.

\bibitem{teddy_eco_alexi}
A.~Alexi, T.~Lazebnik, and L.~Shami.
\newblock Microfounded tax revenue forecast model with heterogeneous population and genetic algorithm approach.
\newblock {\em Computational Economics}, 2023.

\bibitem{ga_initial}
T.~Grossman and A.~Wool.
\newblock Computational experience with approximation algorithms for the set covering problem.
\newblock {\em European Journal of Operational Research}, 101(1):81--92, 1997.

\bibitem{ga_intro}
J.~H. Holland.
\newblock Genetic algorithms.
\newblock {\em Scientific American}, 267(1):66--73, 1992.

\bibitem{cross_over}
L.~Davis.
\newblock Applying adaptive algorithms to epistatic domains.
\newblock {\em Proceedings of the international joint conference on artificial intelligence}, pages 162--164, 1985.

\bibitem{selection_operator}
Z.~W. Bo, L.~Z. Hua, and Z.~G. Yu.
\newblock Optimization of process route by genetic algorithms.
\newblock {\em Robotics and Computer-Integrated Manufacturing}, 22:180--188, 2006.

\bibitem{tskm_intro}
B.~Cai, G.~Huang, N.~Samadiani, G.~Li, and C-H. Chi.
\newblock Efficient time series clustering by minimizing dynamic time warping utilization.
\newblock {\em IEEE Access}, 9:46589--46599, 2021.

\bibitem{elbow}
M.~A. Syakur, B.~K. Khotimah, E.~M.~S. Rochman, and B.~D. Satoto.
\newblock Integration k-means clustering method and elbow method for identification of the best customer profile cluster.
\newblock {\em IOP Conference Series: Materials Science and Engineering}, 336(1):012017, apr 2018.

\bibitem{xmeans}
S.~Soheily-Khah, A.~Douzal-Chouakria, and E.~Gaussier.
\newblock Generalized k-means-based clustering for temporal data under weighted and kernel time warp.
\newblock {\em Pattern Recognition Letters}, 75:63--69, 2016.

\bibitem{rnn}
J.~Zhang and K.~F. Man.
\newblock Time series prediction using rnn in multi-dimension embedding phase space.
\newblock In {\em SMC'98 Conference Proceedings. 1998 IEEE International Conference on Systems, Man, and Cybernetics (Cat. No.98CH36218)}, volume~2, pages 1868--1873, 1998.

\bibitem{adam_optimizator}
D.~P. Kingma and J.~Ba.
\newblock Adam: A method for stochastic optimization.
\newblock {\em arXiv}, 2017.

\bibitem{markov_book}
N.~Privault.
\newblock {\em Understanding Markov Chains}.
\newblock Springer Singapore, 2018.

\bibitem{markov_astronomy}
S.~Sharma.
\newblock Markov chain monte carlo methods for bayesian data analysis in astronomy.
\newblock {\em Annual Review of Astronomy and Astrophysics}, 55(1):213--259, 2017.

\bibitem{autoencoder}
D.~Bank, N.~Koenigstein, and R.~Giryes.
\newblock {\em Autoencoders}, pages 353--374.
\newblock Springer International Publishing, Cham, 2023.

\bibitem{teddy_cancer}
T.~Lazebnik and L/~Simon-Keren.
\newblock Cancer-inspired genomics mapper model for the generation of synthetic dna sequences with desired genomics signatures.
\newblock {\em Computers in Biology and Medicine}, 164:107221, 2023.

\bibitem{tpot}
R.~S. Olson and J.~H. Moore.
\newblock Tpot: A tree-based pipeline optimization tool for automating machine learning.
\newblock In {\em Workshop on Automatic Machine Learning}, pages 66--74. PMLR, 2016.

\bibitem{teddy_bio_inspired_auto_ml}
T.~Lazebnik, T.~Fleischer, and A.~Yaniv-Rosenfeld.
\newblock Benchmarking biologically-inspired automatic machine learning for economic tasks.
\newblock {\em Sustainability}, 15(14), 2023.

\bibitem{drl_teddy}
T.~Lazebnik.
\newblock Data-driven hospitals staff and resources allocation using agent-based simulation and deep reinforcement learning.
\newblock {\em Engineering Applications of Artificial Intelligence}, 126:106783, 2023.

\bibitem{drl_1}
S.~Witty, J.~K. Lee, E.~Tosch, A.~Atrey, K.~Clary, M.~L. Littman, and D.~Jensen.
\newblock Measuring and characterizing generalization in deep reinforcement learning.
\newblock {\em Applied AI Letters}, 2(4):e45, 2021.

\bibitem{drl_2}
C.~Zhang, W.~Song, Z.~Cao, J.~Zhang, P.~S. Tan, and C.~Xu.
\newblock Learning to dispatch for job shop scheduling via deep reinforcement learning.
\newblock In {\em 34th Conference on Neural Information Processing Systems}, 2020.

\bibitem{syntheye}
Y.~A. Veturi, W.~Woof, T.~Lazebnik, I.~Moghul, P.~Woodward-Court, S.~K. Wagner, T.~a. Cabral~de Guimarães, D.~Varela, M, B.~Liefers, P.~J. Patel, S.~Beck, A.~R. Webster, O.~Mahroo, P.~A. Keane, M.~Michaelides, K.~Balaskas, and N.~Ponikos.
\newblock Syntheye: Investigating the impact of synthetic data on artificial intelligence-assisted gene diagnosis of inherited retinal disease.
\newblock {\em Ophthalmology Science}, 3:100258, 2023.

\bibitem{rl_review}
K.~Arulkumaran, M.~P. Deisenroth, M.~Brundage, and A.~A. Bharath.
\newblock Deep reinforcement learning: A brief survey.
\newblock {\em IEEE Signal Processing Magazine}, 34(6):26--38, 2017.

\bibitem{ppo}
J.~Schulman, F.~Wolski, P.~Dhariwal, A.~Radford, and O.~Klimov.
\newblock Proximal policy optimization algorithm.
\newblock {\em arXiv}, 2017.

\bibitem{teddy_pandemic_management}
T.~Lazebnik, S.~Bunimovich-Mendrazitsky, and L.~Shami.
\newblock Pandemic management by a spatio–temporal mathematical model.
\newblock {\em International Journal of Nonlinear Sciences and Numerical Simulation}, 107(4):106176, 2021.

\bibitem{teddy_multi_strain}
T.~Lazebnik and S.~Bunimovich-Mendrazitsky.
\newblock Generic approach for mathematical model of multi-strain pandemics.
\newblock {\em Plos One}, 17(4):e0260683, 2022.

\bibitem{hyperparamters_are_hard}
L.~Yang and A.~Shami.
\newblock On hyperparameter optimization of machine learning algorithms: Theory and practice.
\newblock {\em Neurocomputing}, 415:295--316, 2020.

\bibitem{airport_case}
T.~Harweg, M.~Wagner, and F.~Weichert.
\newblock Agent-based simulation for infectious disease modelling over a period of multiple days, with application to an airport scenario.
\newblock {\em International Journal of Environmental Research and Public Health}, 20(1), 2023.

\bibitem{restaurant_case}
K.~S. Kwon, J.~I. Park, Y.~J. Park, D.~M. Jung, K.~W. Ryu, and J.~H. Lee.
\newblock Evidence of long-distance droplet transmission of sars-cov-2 by direct air flow in a restaurant in korea.
\newblock {\em J Korean Med Sci}, 35(46):e415, 2020.

\bibitem{bus_case}
Y.~Shen, C.~Li, H.~Dong, Z.~Wang, L.~Martinez, Z.~Sun, A.~Handel, Z.~Chen, E.~Chen, M.~H. Ebell, F.~Wang, B.~Yi, H.~Wang, X.~Wang, A.~Wang, B.~Chen, Y.~Qi, L.~Liang, Y.~Li, F.~Ling, J.~Chen, and G.~Xu.
\newblock Community outbreak investigation of sars-cov-2 transmission among bus riders in eastern china.
\newblock {\em JAMA Internal Medicine}, 180(12):1665--1671, 2020.

\bibitem{teddy_chaos}
L.~Shami and T.~Lazebnik.
\newblock Economic aspects of the detection of new strains in a multi-strain epidemiological–mathematical model.
\newblock {\em Chaos, Solitons \& Fractals}, 165:112823, 2022.

\bibitem{end_1}
V.~L.~S. Silva, C.~E. Heaney, Y.~Li, and C.~C. Pain.
\newblock Data assimilation predictive gan (da-predgan) applied to a spatio-temporal compartmental model in epidemiology.
\newblock {\em Journal of Scientific Computing}, 94(25), 2023.

\bibitem{end_2}
X.~Yu, M.~Fang, J.~Yu, L.~Cheng, S.~Ding, and Z.~Kou.
\newblock Epidemiological characteristics and spatio-temporal analysis of brucellosis in shandong province, 2015–2021.
\newblock {\em BMC Infectious Diseases}, 23(669), 2023.

\bibitem{end_3}
E.~Orozco-Acosta, A.~Adin, and M.~D. Ugarte.
\newblock Big problems in spatio-temporal disease mapping: Methods and software.
\newblock {\em Cpmputer Methods and PRograms in Biomedicine}, 231(107403), 2023.

\bibitem{end_4}
S.~Yin, J.~Wu, and P.~Song.
\newblock Optimal control by deep learning techniques and its applications on epidemic models.
\newblock {\em Journal of Mathematical Biology}, 86(36), 2023.

\bibitem{end_5}
I~McDowell.
\newblock {\em Explanation and Causal Models for Social Epidemiology}, pages 37--88.
\newblock Springer International Publishing, 2023.

\end{thebibliography}
\bibliographystyle{unsrt}

\section*{Appendix}

\subsection*{Temporal component for Epidemiological models}
In this sub-section, the formal ordinary differential equation (ODE), as well as the agent-based simulation (ABS) representation are provided for the three models used in the experiments. 

\subsubsection*{SIR}
For the SIR model, the ODE representation takes the form:
\begin{equation}
    \begin{array}{l}
        \frac{dS(t)}{dt} = - \beta S(t)I(t), \\ \\
        \frac{dI(t)}{dt} = \beta S(t)I(t) - \gamma I(t), \\ \\
        \frac{dR(t)}{dt} = \gamma I(t), 
    \end{array}
    \label{eq:appendix_sir}
\end{equation}
where \(S(t), I(t)\), and \(R(t)\) are the number of susceptible, infected, and recovered agents in the population, \(\beta\) is the average infection rate, and \(\gamma\) is the average recovery rate. 

For the OBS representation, let us assume each agent has an inner clock, \(\theta \in \mathbb{N}\), that indicates the number of steps in time passed since the agent's epidemiological state changed last time. Moreover, the epidemiological state (\(\xi\)) belongs to \(\xi \in \{S, I, R\}\). 

\begin{equation}
    \begin{cases}
        \xi = S \times \xi =  I \rightarrow \xi =  I \times \xi =  I, & \beta \\
        \xi = I, \theta = \gamma \rightarrow \xi = R \wedge \theta = 0, & 1 
    \end{cases}.
\label{eq:agents_infection}
\end{equation}
Any other case that is not specifically mentioned in Eq. (\ref{eq:agents_infection}), is the identity function with probability \(1\). Moreover, for all transactions, \(\theta = 0\).

\subsubsection*{SEIRD with two age groups}
For the SEIRD with two age groups model, the ODE representation takes the form:

\begin{equation}
    \begin{array}{l}
    \frac{dS_c(t)}{dt} = - \big ( \beta_{cc}^sI_c^s(t) + \beta_{cc}^aI_c^a(t) + \beta_{ca}^sI_a^s(t) + \beta_{ca}^aI_a^a(t) \big )  S_c(t) , \\\\
    
    \frac{dS_a(t)}{dt} = - \big (\beta_{ac}^sI_c^s(t) + \beta_{ac}^aI_c^a(t) + \beta_{aa}^sI_a^s(t) + \beta_{aa}^aI_a^a(t) \big )  S_a(t), \\\\
    
    \frac{dI_c^s(t)}{dt} =  (1 - \psi_c) \big ( \beta_{cc}^sI_c^s(t) + \beta_{cc}^aI_c^a(t) + \beta_{ca}^sI_a^s(t) + \beta_{ca}^aI_a^a(t) \big ) S_c(t) - \gamma_c I_c^s(t) , \\\\
    
    \frac{dI_c^a(t)}{dt} =  \psi_c  \big ( \beta_{cc}^sI_c^s(t) + \beta_{cc}^aI_c^a(t) + \beta_{ca}^sI_a^s(t) + \beta_{ca}^aI_a^a(t) \big )  S_c(t) - \gamma_c I_c(t), \\\\
    
    \frac{dI_a^s(t)}{dt} =  \psi_a  \big ( \beta_{ac}^sI_c^s(t) + \beta_{ac}^aI_c^a(t) + \beta_{aa}^sI_a^s(t) + \beta_{aa}^aI_a^a(t) \big ) S_a(t) - \gamma_a  I_a(t) , \\\\
    
    \frac{dI_a^a(t)}{dt} =  (1 - \psi_a) \big ( \beta_{ac}^sI_c^s(t) + \beta_{ac}^aI_c^a(t) + \beta_{aa}^sI_a^s(t) + \beta_{aa}^aI_a^a(t) \big )  S_a(t) - \gamma_a  I_a^a(t), \\\\
    
    \frac{dR_c(t)}{dt} = \gamma_c \rho_c (I_c^s(t) + I_c^a(t)), \\\\
    
    \frac{dR_a(t)}{dt} = \gamma_a \rho_a (I_a^s(t) + I_a^a(t)), \\\\
    
        \frac{dD_c(t)}{dt} = \gamma_c (1 - \rho_c) (I_c^s(t) + I_c^a(t)), \\\\
        
        \frac{dD_a(t)}{dt} = \gamma_a  (1 - \rho_a)  (I_a^s(t) + I_a^a(t)),
        \label{eq:final}
    \end{array}
\end{equation}
where \(S(t), I^a(t), I^s(t), R(t)\), and \(D(t)\) are the number of susceptible, asymptomatic infected, symptomatic infected,  recovered, and dead agents in the population, \(\beta\) is the average infection rate, \(\gamma\) is the average recovery rate, \(\rho\) is the recovery probability. 

For the OBS representation, let us assume each agent has an inner clock, \(\theta \in \mathbb{N}\), that indicates the number of steps in time passed since the agent's epidemiological state changed last time and an age group, \(\zeta \in \{c, a\}\). Moreover, the epidemiological state (\(\xi\)) belongs to \(\xi \in \{S, I^a, I^s, R, D\}\). 

\begin{equation}
    \begin{cases}
        \xi = S, \zeta = c \times \xi = I^a, \zeta = c  \rightarrow \xi =  I^a, \zeta = c \times \xi =  I^a, \zeta = c, & \psi_c \beta^a_{cc} \\
        
        \xi = S, \zeta = c \times \xi = I^a, \zeta = c  \rightarrow \xi =  I^s, \zeta = c \times \xi =  I^a, \zeta = c, & (1 - \psi_c) \beta^a_{cc} \\
        
        \xi = S, \zeta = c \times \xi = I^s, \zeta = c  \rightarrow \xi =  I^a, \zeta = c \times \xi =  I^s, \zeta = c, & \psi_c \beta^s_{cc} \\
        
        \xi = S, \zeta = c \times \xi = I^s, \zeta = c  \rightarrow \xi =  I^a, \zeta = c \times \xi =  I^s, \zeta = c, & (1 - \psi_c) \beta^s_{cc} \\
        
        \xi = S, \zeta = c \times \xi = I^a, \zeta = a  \rightarrow \xi =  I^a, \zeta = c \times \xi =  I^a, \zeta = a, & \psi_c \beta^a_{ca} \\
        
        \xi = S, \zeta = c \times \xi = I^a, \zeta = a \rightarrow \xi =  I^s, \zeta = c \times \xi =  I^a, \zeta = a, & (1 - \psi_c) \beta^a_{ca} \\
        
        \xi = S, \zeta = c \times \xi = I^s, \zeta = a  \rightarrow \xi =  I^a, \zeta = c \times \xi =  I^s, \zeta = a, & \psi_c \beta^s_{ca} \\
        
        \xi = S, \zeta = c \times \xi = I^s, \zeta = a  \rightarrow \xi =  I^a, \zeta = c \times \xi =  I^s, \zeta = a, & (1 - \psi_c) \beta^s_{ca} \\
        
        \xi = S, \zeta = a \times \xi = I^a, \zeta = c  \rightarrow \xi =  I^a, \zeta = a \times \xi =  I^a, \zeta = c, & \psi_a \beta^a_{ac} \\
        
        \xi = S, \zeta = a \times \xi = I^a, \zeta = c \rightarrow \xi =  I^s, \zeta = a \times \xi =  I^a, \zeta = c, & (1 - \psi_a) \beta^a_{ac} \\
        
        \xi = S, \zeta = a \times \xi = I^s, \zeta = c  \rightarrow \xi =  I^a, \zeta = a \times \xi =  I^s, \zeta = c, & \psi_a \beta^s_{ac} \\
        
        \xi = S, \zeta = a \times \xi = I^s, \zeta = c  \rightarrow \xi =  I^a, \zeta = a \times \xi =  I^s, \zeta = c, & (1 - \psi_a) \beta^s_{ac} \\
        
        \xi = S, \zeta = a \times \xi = I^a, \zeta = a  \rightarrow \xi = I^a, \zeta = a \times \xi =  I^a, \zeta = a, & \psi_a \beta^a_{aa} \\
        
        \xi = S, \zeta = a \times \xi = I^a, \zeta = a \rightarrow \xi = I^s, \zeta = a \times \xi =  I^a, \zeta = a, & (1 - \psi_a) \beta^a_{aa} \\
        
        \xi = S, \zeta = a \times \xi = I^s, \zeta = a  \rightarrow \xi =  I^a, \zeta = a \times \xi =  I^s, \zeta = a, & \psi_a \beta^s_{aa} \\
        
        \xi = S, \zeta = a \times \xi = I^s, \zeta = a  \rightarrow \xi =  I^a, \zeta = a \times \xi =  I^s, \zeta = a, & (1 - \psi_a) \beta^s_{aa} \\
        
        \xi = I^s, \zeta = c, \theta = \gamma_c \rightarrow \xi = R & \rho_c  \\ 
        
        \xi = I^s, \zeta = a, \theta = \gamma_a \rightarrow \xi = R & \rho_a  \\
        
        \xi = I^s, \zeta = c, \theta = \gamma_c \rightarrow \xi = D & 1 - \rho_c  \\ 
        
        \xi = I^s, \zeta = a, \theta = \gamma_a \rightarrow \xi = D & 1 - \rho_a  \\
        
        \xi = I^a, \zeta = c, \theta = \gamma_c \rightarrow \xi = R & 1  \\ 
        
        \xi = I^a, \zeta = a, \theta = \gamma_a \rightarrow \xi = R & 1  \\
        
    \end{cases}.
\label{eq:agents_infection}
\end{equation}
Any other case that is not specifically mentioned in Eq. (\ref{eq:agents_infection}), is the identical function with chance \(1\). Moreover, for all transactions, \(\theta = 0\).

\subsubsection*{Two-strain SIR}
For the two-strain SIR model, the ODE representation takes the form:

\begin{equation}
    \begin{array}{ll}
    \frac{dR_ \emptyset I_1(t)}{dt} = \beta_{\emptyset,1} (R_{\emptyset}I_{1}(t) + R_{\{ 2 \}}I_{1}(t))  R_{\emptyset}(t)  -   \gamma_{\emptyset,1} R_{\emptyset}I_{1}(t)  , \\\\ 
    
    \frac{dR_{\{ 2\}} I_1(t)}{dt} = \beta_{\{ 2 \},1} (R_{\{ 2 \}}I_{1}(t) + R_{\emptyset}I_{1}(t)) R_{\{ 2 \}}(t)  -   \gamma_{\{ 2 \},1} R_{\{ 2 \}}I_{1}(t)  , \\\\
    
    \frac{dR_ \emptyset I_2(t)}{dt} = \beta_{\emptyset,2} (R_{\emptyset}I_{2}(t) + R_{\{ 1 \}}I_{2}(t)) R_{\emptyset}(t)  -   \gamma_{\emptyset,2} R_{\emptyset}I_{2}(t)  , \\\\
    
    \frac{dR_{\{ 1\}} I_2(t)}{dt} = \beta_{\{ 1 \},2} (R_{\{ 1 \}}I_{2}(t) + R_{\emptyset}I_{2}(t)) R_{\{ 1 \}}(t)  -   \gamma_{\{ 1 \},2} R_{\{ 1 \}}I_{2}(t)  , \\\\
    
    \frac{dR_\emptyset(t)}{dt} = - R_\emptyset(t) \big ( \beta_{\emptyset,1} (R_{\emptyset}I_{1}(t) + R_{\{ 2 \}}I_{1}(t)) + \beta_{\emptyset,2} (R_{\emptyset}I_{2}(t) + R_{\{ 1 \}}I_{2}(t))  \big  ), \\\\
    
    \frac{dR_{\{1\}}(t)}{dt} = \gamma_{\emptyset,1} \rho_{\emptyset,1} R_{\emptyset}I_{1}(t)  -  \beta_{\{ 1 \},2} (R_{\{ 1 \}}I_{2}(t) + R_{\emptyset}I_{2}(t)) R_{\{ 1 \}}(t) , \\\\
        
    \frac{dR_{\{2\}}(t)}{dt} = \gamma_{\emptyset,2} \rho_{\emptyset,2} R_{\emptyset}I_{2}(t)  -  \beta_{\{ 2 \},1} (R_{\{ 2 \}}I_{1}(t) + R_{\emptyset}I_{1}(t)) R_{\{ 2 \}}(t) , \\\\
    
    \frac{dR_{\{1,2\}}(t)}{dt} = \gamma_{\{2\},1} \rho_{\{2\},1} R_{\{2\}}I_{1}(t) + \gamma_{\{1\},2} \rho_{\{1\},2} R_{\{1\}}I_{2}(t), \\\\
        
    \begin{split}
    \frac{dD(t)}{dt} = \gamma_{\emptyset,1} (1 - \rho_{\emptyset,1}) R_{\emptyset}I_{1}(t) + \gamma_{\{2\},1} (1 - \rho_{\{2\},1}) R_{\{2\}}I_{1}(t) \\ + \gamma_{\emptyset,2} (1 - \rho_{\emptyset,2}) R_{\emptyset}I_{2}(t) + \gamma_{\{1\},2} (1 - \rho_{\{1\},2}) R_{\{1\}}I_{2}(t) . \\\\
       \end{split}
\end{array}
\label{eq:two_mulation_case}
\end{equation}
where \(R_\emptyset(t), R_{\{1\}}, R_{\{2\}}, R_{\{1, 2\}}, R_\emptyset I_1(t), R_\emptyset I_2(t), R_{\{1\}} I_2, R_{\{2\}} I_1 \), and \(D(t)\) are the number of recovered from strains \(\emptyset, \{1\}, \{2\}, \{1,2\}\), infected with the first and second strains, and dead agents in the population, \(\beta\) is the average infection rate, \(\gamma\) is the average recovery rate, \(\psi\) is the recovery probability. 

For the OBS representation, let us assume each agent has an inner clock, \(\theta \in \mathbb{N}\), that indicates the number of steps in time passed since the agent's epidemiological state changed last time and an age group. Moreover, the epidemiological state (\(\xi\)) belongs to \(\xi \in \{R_\emptyset(t), R_{\{1\}}, R_{\{2\}}, R_{\{1, 2\}}, R_\emptyset I_1(t), R_\emptyset I_2(t), R_{\{1\}} I_2, R_{\{2\}} I_1, D\}\). 

\begin{equation}
    \begin{cases}
        \xi = \{\emptyset, 1\} \times \theta =  \gamma_{\emptyset, 1} \rightarrow \xi =  D, & 1 - \rho_{\emptyset, 1} \\
        \xi = \{\emptyset, 2\} \times \theta =  \gamma_{\emptyset, 2} \rightarrow \xi =  D, & 1 - \rho_{\emptyset, 2} \\
        \xi = \{\{1\}, 2\} \times \theta =  \gamma_{\{1\}, 2} \rightarrow \xi =  D, & 1 - \rho_{\{1\}, 2} \\
        \xi = \{\{2\}, 1\} \times \theta =  \gamma_{\{2\}, 1} \rightarrow \xi =  D, & 1 - \rho_{\{2\}, 1} \\
        \xi = \{\emptyset, 1\} \times \theta =  \gamma_{\emptyset, 1} \rightarrow \xi = \{\{1\}, \emptyset\}, & \rho_{\emptyset, 1} \\
        \xi = \{\emptyset, 2\} \times \theta =  \gamma_{\emptyset, 2} \rightarrow \xi =  \{\{2\}, \emptyset\}, & \rho_{\emptyset, 2} \\
        \xi = \{\{1\}, 2\} \times \theta =  \gamma_{\{1\}, 2} \rightarrow \xi = \{\{1, 2\}, \emptyset\}, & \rho_{\{1\}, 2} \\
        \xi = \{\{2\}, 1\} \times \theta =  \gamma_{\{2\}, 1} \rightarrow \xi = \{\{1, 2\}, \emptyset\}, & \rho_{\{2\}, 1} \\
        \xi = \{\emptyset, \emptyset\} \times \xi = \{\emptyset, 1\} \rightarrow \xi = \{\emptyset, 1\} \times \xi = \{\emptyset, 1\}, & \beta_{\emptyset, 1} \\
        \xi = \{\emptyset, \emptyset\} \times \xi = \{\emptyset, 2\} \rightarrow \xi = \{\emptyset, 2\} \times \xi = \{\emptyset, 2\}, & \beta_{\emptyset, 2} \\
        
        \xi = \{\{1\}, \emptyset\} \times \xi = \{\emptyset, 2\} \rightarrow \xi = \{\{1\}, 2\} \times \xi = \{\emptyset, 2\}, & \beta_{\{1\}, 2} \\ 
        
        \xi = \{\{2\}, \emptyset\} \times \xi = \{\emptyset, 1\} \rightarrow \xi = \{\{1\}, 1\} \times \xi = \{\emptyset, 1\}, & \beta_{\{2\}, 1} \\ 
    \end{cases}.
\label{eq:agents_infection}
\end{equation}

Any other case that is not specifically mentioned in Eq. (\ref{eq:agents_infection}), is the identical function with chance \(1\). Moreover, for all transactions, \(\theta = 0\).

\end{document}